\documentclass[letterpaper,american,american,manuscript,showpacs]{revtex4}
\usepackage[T1]{fontenc}
\usepackage[latin9]{inputenc}
\usepackage{color}
\usepackage{amsmath}
\usepackage{amssymb}
\usepackage{graphicx}

\makeatletter

\@ifundefined{textcolor}{}
{%
 \definecolor{BLACK}{gray}{0}
 \definecolor{WHITE}{gray}{1}
 \definecolor{RED}{rgb}{1,0,0}
 \definecolor{GREEN}{rgb}{0,1,0}
 \definecolor{BLUE}{rgb}{0,0,1}
 \definecolor{CYAN}{cmyk}{1,0,0,0}
 \definecolor{MAGENTA}{cmyk}{0,1,0,0}
 \definecolor{YELLOW}{cmyk}{0,0,1,0}
}

\@ifundefined{definecolor}
 {\usepackage{color}}{}

\@ifundefined{definecolor}
 {\@ifundefined{definecolor}
 {\usepackage{color}}{}
}{}

\usepackage{babel}

\makeatother

\usepackage{babel}

\makeatother

\usepackage{babel}
\begin{document}

\title{Guiding synchrony through random networks}

\author{Sven Jahnke$^{1-3}$, Marc Timme$^{1-3}$, Raoul-Martin Memmesheimer$^{4}$}

\affiliation{$^{1}$Network Dynamics Group, Max Planck Institute for Dynamics
\& Self-Organization (MPIDS), }

\affiliation{$^{2}$Bernstein Center for Computational Neuroscience (BCCN), 37073
Göttingen, Germany,}

\affiliation{$^{3}$Fakultät für Physik, Georg-August-Universität Göttingen,}

\affiliation{$^{4}$Donders Institute, Department for Neuroinformatics, Radboud
University, Nijmegen, Netherlands.}

\date{Mon June 18 2012}
\begin{abstract}
Sparse random networks contain structures that can be considered as
diluted feed-forward networks. Modeling of cortical circuits has shown
that feed-forward structures, if strongly pronounced compared to the
embedding random network, enable reliable signal transmission by propagating
localized (sub-network) synchrony. This assumed prominence, however,
is not experimentally observed in local cortical circuits. Here we
show that nonlinear dendritic interactions as discovered in recent
single neuron experiments, naturally enable guided synchrony propagation
already in random recurrent neural networks exhibiting mildly enhanced,
biologically plausible sub-structures. 
\end{abstract}


\maketitle

\section{Introduction}

Cortical neural networks generate a ground state of highly irregular
spiking activity whose dynamics is sensitive to small perturbations
such as missing or additional spikes \cite{Vreeswijk1996,Brunel2000,JMT08,LRBHL10}.
A robust, reliable transmission of information in the presence of
such perturbations and noise is nonetheless assumed to be essential
for neural computation. It has been hypothesized that this might be
achieved by propagation of pulses of synchronous spikes along feed-forward
chains \cite{SynChains}. In current models, functionally relevant
chains require a dense connectivity between the neuronal layers of
the network \cite{EmbSyns} or strongly enhanced synapses and specifically
modified response properties of neurons within the chain \cite{Vogels2005}.
Such highly distinguished large-scale structures however are not observed
experimentally.

Can less structured networks also guide synchrony? Recently, single
neuron experiments have revealed a mechanism that nonlinearly promotes
synchronous inputs. Upon synchronous dendritic stimulation, neurons
are capable of generating fast dendritic spikes. In the soma, these
induce rapid, strong depolarizations \cite{DSpikes} that are nonlinearly
enhanced compared to depolarizations expected from linear summation
of single inputs. If the dendritic spike induces an action potential in the soma,
the latter occurs at a fixed time after the stimulation, with sub-millisecond
precision. Other experiments have found slow dendritic spikes which
are comparably insensitive to input synchrony \cite{Hausser2000}.
These slow dendritic spikes endow single neurons with computational
capabilities comparable to multi-layered feed-forward networks of
simple rate neurons \cite{slowD}. Furthermore, they provide a possible
mechanism underlying neural bursting and its propagation, which have
been shown to enhance reliability and temporal precision of signal
propagation \cite{bursts,LJF2010}. The impact of fast dendritic spikes
that induce non-additive coupling, on collective circuit dynamics
has not been systematically investigated so far in a general setting.

In this article, we show that and how fast dendritic nonlinearities
may support guided synchrony propagation in neural circuits. First,
we develop an analytical approach to describe such propagation in
linearly and nonlinearly coupled networks. In particular, we derive
an expression for the critical connectivity above which propagation
occurs and for the size of the propagating pulse. We quantify how
dendritic nonlinearities compensate for dense anatomical connections
and thereby promote propagation of synchrony. Finally, using large-scale
simulations of more detailed recurrent network models, we show that
feed-forward networks that occur naturally as part of random circuits
enable persistent guided synchrony propagation due to dendritic nonlinearities.

\section{Models and Methods}

\subsection{Analytically tractable model}

\textit{Model with linear summation of inputs. }As a basis model we
consider networks of conventional leaky integrate-and-fire neurons
that interact by sending and receiving spikes via directed connections.
The membrane potential $V_{l}$ of a neuron $l$ satisfies
\begin{equation}
\dot{V}_{l}(t)=-\gamma_{l}V_{l}(t)+I_{l}\left(t\right),\label{eq:model}
\end{equation}
where $\gamma_{l}$ is the inverse membrane time constant and $I_{l}(t)$
is the total input current at time $t$. In addition to inputs from
the network, the neurons receive excitatory and inhibitory random
inputs which emulate an embedding network, i.e.

\begin{equation}
I_{l}(t)=I_{l}^{0}+I_{l}^{\text{ext,ex}}(t)+I_{l}^{\text{ext,in}}(t)+I_{l}^{\text{net}}(t),
\end{equation}
where $I_{l}^{0}$ is a constant input current modeling slow external
(from outside the chain) and internal (from the chain) currents, $I_{l}^{\text{ext,ex}}(t)$
and $I_{l}^{\text{ext,in}}(t)$ are the contributions due to arriving
external excitatory and inhibitory spikes (that are modeled as independent
random (Poissonian) spike trains with rate $\nu^{\text{ext,ex}}$
and $\nu^{\text{ext,in}}$ respectively) and $I_{l}^{\text{net}}(t)$
are the contributions originating from spikes of neurons of the network.
In the absence of any spiking activity, the membrane potential will
exponentially converge towards its asymptotic value $V_{l}^{\infty}:=I_{l}^{0}/\gamma_{l}$.
When the neuron's membrane potential reaches or exceeds its threshold
$\Theta_{l}$, its membrane potential is reset to $V_{l}^{\text{reset}}$
and a spike is emitted, which arrives at the postsynaptic neuron $j$
after a delay time $\tau_{jl}$. For a refractory period $t_{l}^{\text{ref}}$
after the reset, all incoming spikes to neuron $l$ are ignored and
the membrane potential is kept at $V_{l}^{\text{reset}}$.

We model the fast rise of the membrane potential upon the arrival
of a presynaptic spike by an instantaneous jump, such that the contributions
of the arriving external spikes to the total input current are given
by
\begin{eqnarray}
I_{l}^{\text{ext,ex}}(t) & = & \sum_{k\in\mathbb{Z}}\epsilon^{\text{ext,ex}}\cdot\delta\left(t-t_{l,k}^{\text{ext,ex}}\right),\\
I_{l}^{\text{ext,in}}(t) & = & \sum_{k\in\mathbb{Z}}\epsilon^{\text{ext,in}}\cdot\delta\left(t-t_{l,k}^{\text{ext,in}}\right),
\end{eqnarray}
where $t_{l,k}^{\text{ext,ex}}$ ($t_{l,k}^{\text{ext,in}}$) are
the arrival times of the $k$th excitatory (inhibitory) external spike
at neuron $l$, $\epsilon^{\text{ext,ex}}>0$ or $\epsilon^{\text{ext,in}}<0$
are the strengths of single external spikes and $\delta(\cdot)$ is
the Dirac $\delta$-distribution. Analogously the contribution of
spikes received from neurons of the network is given by
\begin{equation}
I_{l}^{\text{net}}\left(t\right)=\sum_{j}\sum_{k}\epsilon_{lj}\delta\left(t-t_{j,k}^{f}-\tau_{lj}\right),\label{eq:InetLin}
\end{equation}
where $\epsilon_{lj}$ is the coupling strength from neuron $j$ to
$l$ and $t_{j,k}^{f}$ is the $k$th spike time of neuron $j$.

\textit{Model with nonlinear summation of inputs.} In the above model
without nonlinear dendrites, the strengths of synchronous inputs are
summed up linearly (cf.~Eq.~\eqref{eq:InetLin}). We incorporate
nonlinear dendrites by modulating this sum for excitatory inputs by
a nonlinear function $\sigma$, that can be directly read off from
experimental results \cite{DSpikes}: $\sigma$ equals the identity
for small excitatory input, increases steeply when the input exceeds
a threshold $\Theta_{b}$ and saturates for larger inputs. We define
the dendritic modulation function as 
\begin{equation}
\sigma\left(\epsilon\right)=\begin{cases}
\epsilon & \text{for }\epsilon\leq\Theta_{b}\\
\kappa & \text{otherwise}
\end{cases}.\label{eq:sigma}
\end{equation}
For simplicity, we consider only exactly simultaneous spikes as synchronous.
Accordingly, conduction delays are chosen homogeneously, $\tau_{ij}\equiv\tau$,
so that synchronous presynaptic spiking can be amplified. In this
scenario the detection of synchronous events is straightforward. However,
systems with heterogeneous delays and finite dendritic integration
window exhibit qualitatively the same phenomena \cite{inPrep}. The
contribution of spikes received from the network is then given by
\begin{eqnarray}
I_{l}^{\text{net}}(t) & = & \sum_{t^{f}}\left[\sigma\left(\sum_{j\in M_{ex}\left(t^{f}\right)}\epsilon_{lj}\right)+\sum_{j\in M_{in}\left(t^{f}\right)}\epsilon_{lj}\right]\nonumber \\
 &  & \quad\times\delta\left(t-t^{f}-\tau\right),
\end{eqnarray}
where $t^{f}$ are all firing times in the network. The sets $M_{ex}(t^{f})$
and $M_{in}(t^{f})$ denote the sets of indices of neurons sending
an excitatory and inhibitory spike at time $t^{f}$, respectively.
Networks with linear dendrites can be described by setting $\sigma(\epsilon)=\epsilon$.\\

\subsection{Biologically more detailed model}

\begin{figure*}
\centering{}\includegraphics[width=1\textwidth]{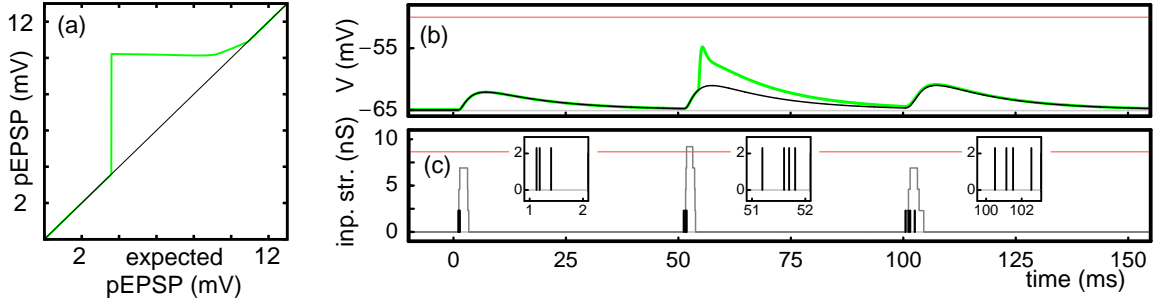}\caption{(color) Example dynamics of a conductance-based leaky integrate-and-fire
neuron with dendritic spike generation (the neuron is initially at
resting membrane potential $V^{\text{rest}}=-65$mV, there are no
external inputs and $I^{0}=0$). Panel (a) shows the pEPSP after a
stimulation versus the expected pEPSP, i.e.~the pEPSP for a neuron
without dendritic spike generation. For inputs corresponding to a
pEPSP larger than $\approx3.8$mV, a dendritic spike is generated
which leads to a higher depolarization than expected from additive
integration. Panel (b) shows the time course of the membrane potential
of a neuron with (green) and without (black) nonlinear dendritic interaction
in response to different excitatory inputs sequences (red horizontal
line: somatic spike threshold). Panel (c) shows the input sequences
(black lines, strength: $g^{\text{ex}}=2.3$nS, closeups given by
insets) and the sum $g_{l,\Delta t}(t)$ of excitatory inputs received
within the dendritic integration window $[t-\Delta t,t]$ (gray),
cf.~Eq.~\eqref{eq:SummedExcInp}. At the first spike arrival around
$t=1$ms, three inputs are received within $\Delta t$ such that $g_{l,\Delta t}(t)$
reaches $6.9$nS. The sum is smaller than the dendritic threshold
$g^{\Theta}=8.65$nS (red horizontal line in c), so no dendritic spike
is generated and there is no difference between the membrane potential
for a neuron with and without a mechanism for dendritic spike generation.
Around $t=50$ms, four spikes arrive within $\Delta t$, $g_{l,\Delta t}(t)$
exceeds the dendritic threshold and a dendritic spike is generated.
Around $t=100$ms four spikes arrive at the neuron, but the temporal
difference between the last and the first spike is slightly larger
than $\Delta t$. Consequently, $g_{l,\Delta t}(t)$ does not exceed
the dendritic threshold and no dendritic spike is initiated.\label{fig:props}}
\end{figure*}

\textit{Conductance based model.} In the last part of the article
we employ a biologically more detailed neuron model to highlight the
generality of our findings on propagation enhancement. The neuron
model is a conductance-based leaky integrate-and-fire neuron that
is augmented by terms introducing the impact of dendritic spikes (see
also \cite{Memmesheimer2010}). The subthreshold dynamics of the membrane
potential $V_{l}$ of neuron $l$ obeys the differential equation\begin{widetext}
\begin{equation}
C_{l}^{\text{m}}\frac{dV_{l}(t)}{dt}=g_{l}^{L}\left(V_{l}^{\text{rest}}-V_{l}(t)\right)+g_{l}^{A}(t)\left(E^{\text{Ex}}-V_{l}(t)\right)+g_{l}^{G}(t)\left(E^{\text{In}}-V_{l}(t)\right)+I_{l}^{\text{DS}}(t)+I_{l}^{0}.\label{eq:CBModel}
\end{equation}
 \end{widetext}Here, $C_{l}^{\text{m}}$ is the membrane capacity,
$g_{l}^{L}$ is the resting conductance and $V_{l}^{\text{rest}}$
is the resting membrane potential, $E^{\text{Ex}}$ and $E^{\text{In}}$
are the reversal potentials, and $g_{l}^{A}(t)$ and $g_{l}^{G}(t)$
are the conductances of excitatory and inhibitory synaptic populations,
respectively. $I_{l}^{\text{DS}}(t)$ models the current pulses caused
by dendritic spikes and $I_{l}^{0}$ is a constant current gathering
slow external and internal currents. The time course of single synaptic
conductances contributing to $g_{l}^{A}(t)$ and $g_{l}^{G}(t)$ is
given by the difference between two exponential functions (e.g.~\cite{DA01}).
Whenever the membrane potential reaches the spike threshold $\Theta_{l}$,
the neuron sends a spike to its postsynaptic neurons, is reset to
$V_{l}^{\text{reset}}$ and becomes refractory for a period $t_{l}^{\text{ref}}$. 

To account for dendritic spike generation, we consider the sum $g_{l,\Delta t}$
of excitatory input strengths (characterized by the coupling strengths)
arriving at an excitatory neuron $l$ within the time window $\Delta t$
for nonlinear dendritic interactions,
\begin{equation}
g_{l,\Delta t}(t)=\sum_{j}\sum_{k}g_{lj}^{\text{max}}\chi_{\left[t,t-\Delta t\right]}(t_{j,k}^{f}+\tau),\label{eq:SummedExcInp}
\end{equation}
 where $\chi_{\left[t,t-\Delta t\right]}$ is the characteristic function
of the interval $[t,t-\Delta t]$, $t_{j,k}^{f}$ is the $k$th firing
time of excitatory neuron $j$ and $\tau$ denotes the synaptic delay.
We denote the peak conductance (coupling strength) for a connection
from neuron $j$ to neuron $l$ by $g_{lj}^{\text{max}}$. If $g_{l,\Delta t}$
exceeds a threshold $g_{\Theta}$, a dendritic spike is initiated
and the dendrite becomes refractory for a time window $t^{\text{DS,ref}}$.
The effect of the dendritic spike is incorporated into the model by
the current pulse that reaches the soma a time $\tau^{\text{DS}}$
thereafter. This current pulse is modeled as the sum of three exponential
functions, 
\begin{equation}
I_{l}^{\text{DS}}(t)=c(g_{\Delta t})\left[-Ae^{-\frac{t}{\tau^{\text{DS,1}}}}+Be^{-\frac{t}{\tau^{\text{DS,2}}}}-Ce^{-\frac{t}{\tau^{\text{DS,3}}}}\right],\label{eq:DSModel}
\end{equation}
 with prefactors $A>0$, $B>0$, $C>0$, decay time constants $\tau^{\text{DS,1}}$,
$\tau^{\text{DS,2}}$, $\tau^{\text{DS,3}}$ and a dimensionless correction
factor $c\left(g_{\Delta t}\right)$, where $g_{\Delta t}$ is the
summed excitatory input at the initiation time of the dendritic spike
as given by Eq.~\eqref{eq:SummedExcInp}. The factor $c\left(g_{\Delta t}\right)$
modulates the pulse strength, ensuring that the peak of the excitatory
postsynaptic potential (pEPSP) reaches the experimentally observed
region of saturation. At very high excitatory inputs, the conventionally
generated depolarization exceeds the level of saturation, and the
pEPSP increases (cf.~Fig.~\eqref{fig:props}a).

\textit{Detection probability. }In the last part of the article we
investigate recurrent networks, where a feed-forward subnetwork consisting
of a certain number of layers (groups) is created by modifying strengths
of existing synaptic connections of the network. To decide whether
propagation of synchrony in recurrent networks is successful, we consider
the signal to noise ratio (SNR): We pick $\omega$ neurons, randomly
selected from the network, to be a first group. After initiation of
synchronous activity in this group, we count the number of spikes
from neurons of the $i^{\text{th}}$ group (for details on how the
$i^{\text{th}}$ group is defined, see section on recurrent networks),
$S_{i}$, within a time window $\left[t_{i}^{\text{exp}}-\frac{t^{w}}{2},t_{i}^{\text{exp}}+\frac{t^{w}}{2}\right]$.
Here $t_{i}^{\text{exp}}$ is the expected time for the synchronous
pulse to reach layer $i$ and $t^{w}$ is the expected width of the
synchronous pulse. We consider all spikes within the time window of
size $t^{w}$ centered at $t_{i}^{\text{exp}}$ as part of the synchronous
pulse. We assume that $t_{i}^{\text{exp}}=t_{1}^{\exp}+\left(i-1\right)\Delta t^{\text{exp}}$,
where $\Delta t^{\text{exp}}$ itself is chosen after simulation such
that $\sum_{i}S_{i}$ becomes maximal,\begin{widetext} 
\begin{equation}
S_{i}=\sum_{k}\sum_{j\in\text{Gr}\left(i\right)}\chi_{\left[t_{i}^{\text{exp}}-\frac{t^{w}}{2},t_{i}^{\exp}+\frac{t^{w}}{2}\right]}\left(t_{j,k}^{f}\right).\label{SUP:signal}
\end{equation}
 \end{widetext}Here, $\text{Gr}\left(i\right)$ are the indices of
neurons of group $i$, $t_{j,k}^{f}$ is the $k^{\text{th }}$ firing
time of neuron $j$ and $\chi$ denotes the characteristic function,
as before.

To determine the noise level of group $i$, we measure the probability
$P_{\Delta t^{\text{obs}},t^{w}}^{i}(k)$ of finding $k$ spikes from
neurons of group $i$ within time windows $t^{w}$ over a control
time interval where no synchronous activity is initiated. The noise
level $N_{i}$ of group $i$ is the minimal value satisfying 
\begin{equation}
\sum_{k=0}^{N_{i}}P_{\Delta t^{\text{obs}},t^{w}}^{i}\left(k\right)\geq a,
\end{equation}
with a constant $a\lesssim1.$

Finally, we denote the propagation of synchrony up to the $i^{\text{th}}$
layer as successful, if the SNR is larger than b, 
\begin{equation}
\text{SNR}_{i}:=\min_{j=1,\dots,i}\left\{ \frac{S_{j}}{N_{j}}\right\} >b,
\end{equation}
where $b\geq1.$ This means in particular that we can distinguish
the background (spontaneous) activity from the signal induced by propagation
of synchrony in all layers $1,\ldots,i$.

\section{Results}

\subsection{Feed-forward chains with linear coupling}

How can diluted feed-forward networks (FFNs) propagate synchrony?
FFNs consist of a sequence of layers, each composed of $\omega$ excitatory
neurons, and forward connections to neurons in the subsequent layer
randomly present with probability $p$. Present connections have strength
$\epsilon$. Synchronous spiking activity is initiated by exciting
neurons of the first layer to spike simultaneously. In the second
layer they excite a certain subgroup of neurons to spike simultaneously
that in turn generates a synchronous input to layer three, etc.

\begin{figure*}
\begin{centering}
\includegraphics[width=1\textwidth]{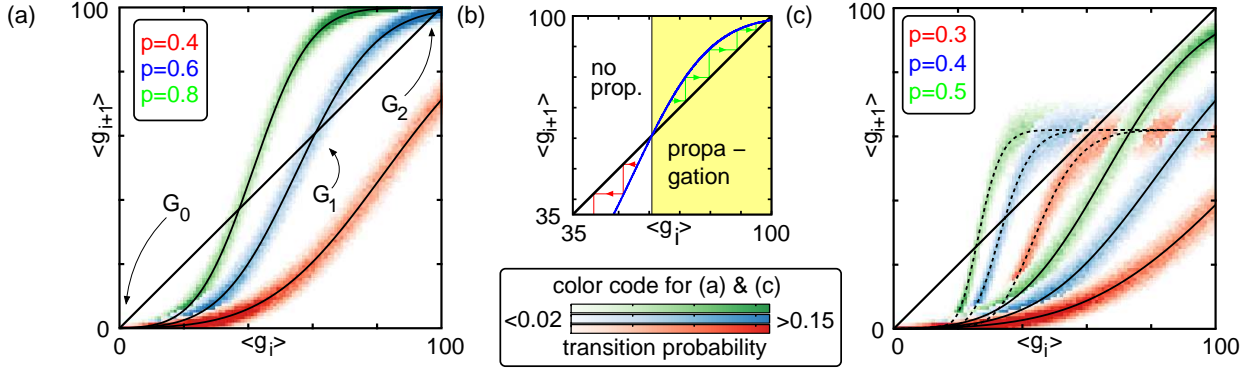} 
\par\end{centering}

\caption{(color online) Emergence of propagation of synchrony. (a) Analytically
derived iterated maps approximating the time evolution of the synchronous
pulse (solid line, cf.~Eq.~\eqref{eq:aveGroupSize}) and transition
probability obtained from network simulations (color code). (b) The
basin of attraction of the stable fixed point $G_{2}$ is illustrated:
Initial pulses within the range $\left(G_{1},\omega\right]$ will
propagate with an average pulse size around $G_{2}$. (c) Iterated
maps for FFNs with linear (solid) and nonlinear dendritic interactions
(dashed). Nonlinear interactions reduce the connectivity required
for propagation and allow for smaller fractions of active neurons.
\label{fig:averagegroup}}
\end{figure*}

To understand the collective dynamics analytically, we consider networks
of leaky integrate-and-fire neurons in the limit of fast synaptic
currents (cf.~Section ``Models and Methods''). In the absence of
synchronous activity, each neuron of the FFN receives a large number
of inputs from an emulated external network and only very few inputs
from the previous layer, such that its dynamics is practically identical
to the ground state of balanced networks. If the connections within
the FFN are weak and/or the connection probability is low, the spontaneous
spiking activity is only weakly influenced by spiking activity of
the FFN. Therefore, we assume that the ground state activity is exclusively
governed by the external inputs, effectively setting couplings within
the chain to $\epsilon_{ij}=0$. The external input is balanced, i.e.~the
mean input is subthreshold and spontaneous spiking is caused by fluctuations
in the input. The network's neurons thus spike in an asynchronous
and irregular manner \cite{Brunel2000,Vreeswijk1996} and the stationary
distribution of membrane potentials $P_{V}(V)$ can be calculated
analytically in diffusion approximation \cite{Brunel2000,Brunel1999Helias2010}.

\begin{equation}
p_{f}(x):=\int_{\Theta-x}^{\Theta}P_{V}(V)dV\label{eq:pfofgi}
\end{equation}
is the probability of finding a neuron's membrane potential in the
interval $\left[\Theta-x,\Theta\right]$. We model the fast rise of
the membrane potential upon the arrival of (possibly nonlinear enhanced)
presynaptic spikes by an instantaneous jump in the membrane potential
(cf.~Section ``Models and Methods''), thus $p_{f}\left(\sigma\left(h\epsilon\right)\right)$
specifies the spiking probability of a single neuron, after receiving
$h$ input spikes of strength $\epsilon$ from the preceding layer.

To assess the propagation of synchrony, we consider the average number
of neurons which are activated in each layer in response to the initial
synchronous pulse (cf.~also~\cite{AverGroup}). When $g_{i}$ neurons
spike synchronously in layer $i$, 
\begin{equation}
p^{\text{sp}}(g_{i}):=\sum_{h=0}^{g_{i}}\binom{g_{i}}{h}p^{h}\left(1-p\right){}^{g_{i}-h}p_{f}\left(\sigma\left(h\epsilon\right)\right)\label{eq:pFofgi}
\end{equation}
 is the probability of spiking of a particular neuron in layer $i+1$,
where the number of simultaneous inputs $h$ is binomially distributed,
$h\sim B(g_{i},p)$. Thus, for layers of size $\omega$, the average
number of neurons spiking in layer $i+1$ is 
\begin{equation}
\left\langle g_{i+1}\right\rangle =\omega p^{\text{sp}}\left(g_{i}\right).\label{eq:aveGroupSize}
\end{equation}
Substituting the average group size $\left\langle g_{i}\right\rangle $
for the actual size $g_{i}$ yields the interpolated map $\left\langle g_{i+1}\right\rangle =\omega p^{\text{sp}}\left(\left\langle g_{i}\right\rangle \right),$
whose fixed points qualitatively determine the propagation of synchronous
activity, cf.~Fig.~\ref{fig:averagegroup}.

The trivial, absorbing fixed point $G_{0}=0$, defining a state of
extinguished activity always exists. For sufficiently small $p$,
$\epsilon$ and $\omega$, this is the only fixed point. With increasing
connectivity and layer size, a pair of fixed points ($G_{1}$, unstable,
and $G_{2}$, stable) appears via a tangent bifurcation. Initial pulses
in the basin of $G_{2}$ (i.e. those larger than $G_{1}$) typically
initiate stable propagation of synchrony with group sizes around $G_{2}$.
For given layer size $\omega$ and connection strength $\epsilon$,
the critical connectivity $p^{\ast}$ for which $G_{1}=G_{2}$ marks
the minimal connectivity that supports stable propagation of synchrony.

To elaborate the influence of nonlinear dendritic interactions, we
derive the critical connectivity for FFNs. The mechanisms underlying
propagation of synchrony are different for networks with and without
nonlinear dendritic interactions and thus require different analytical
approaches to derive $p^{\ast}$. We first consider feed-forward chains
with conventional, linear coupling, i.e.~$\sigma(x)=x$. To obtain
$p^{\ast}$, we first expand $p_{f}(x)$ into a Taylor series up to
first order around the mean of the binomial distribution specifying
the average number $pg_{i}$ of active neurons in each layer, such
that Eq.~\eqref{eq:aveGroupSize} simplifies to
\begin{eqnarray}
\left\langle g_{i+1}\right\rangle  & = & \omega\sum_{h=0}^{g_{i}}\binom{g_{i}}{h}p^{h}\left(1-p\right){}^{g_{i}-h}p_{f}\left(h\epsilon\right)\label{SUP:map}\\
 & \stackrel{\cdot}{=} & \omega\sum_{h=0}^{g_{i}}\binom{g_{i}}{h}p^{h}\left(1-p\right){}^{g_{i}-h}\label{SUP:maplin}\\
 &  & \quad\times\left(p_{f}\left(g_{i}p\epsilon\right)+p_{f}^{\prime}(g_{i}p\epsilon)(h\epsilon-g_{i}p\epsilon)\right)\\
 & = & \omega p_{f}(g_{i}p\epsilon).\label{SUP:AppG1}
\end{eqnarray}
The linear approximation becomes exact in the limit of large layer
sizes $\omega$ and small couplings $\epsilon$, where the product
$\epsilon\omega$ is kept constant. We obtain an interpolated map
from Eq.~\eqref{SUP:AppG1} by replacing $g_{i}$ by its mean value
$\left\langle g_{i}\right\rangle $. At the fixed point $G:=\left\langle g_{i+1}\right\rangle =\left\langle g_{i}\right\rangle $,
the function
\begin{equation}
F(G,p^{\dagger},\omega,\epsilon):=G-\omega p_{f}(p^{\dagger}G\epsilon)=0
\end{equation}
 vanishes. Here, the values $G$ and $p^{\dagger}$ are the average
group size and the connection probability at the fixed point for given
layer size $\omega$ and coupling strength $\epsilon$. Furthermore,
$F$ has a double root at the bifurcation point, so the derivative
\begin{equation}
\frac{\partial F(G,p^{\ast},\omega,\epsilon)}{\partial G}=1-\omega p^{\ast}\epsilon p_{f}^{\prime}(p^{\ast}G\epsilon)=0
\end{equation}
 also vanishes such that the derivative of $p_{f}$ at the bifurcation
point is given by
\begin{equation}
p_{f}^{\prime}\left(p^{\ast}G\epsilon\right)=\frac{1}{\omega p^{\ast}\epsilon}.
\end{equation}
 Combining the above equations, we express the derivatives of $F$
at the bifurcation point by
\begin{eqnarray}
\frac{\partial F(G,p^{\ast},\omega,\epsilon)}{\partial p^{\ast}} & = & -\omega G\epsilon p_{f}^{\prime}(p^{\ast}G\epsilon)=-\frac{G}{p^{\ast}},\\
\frac{\partial F(G,p^{\ast},\omega,\epsilon)}{\partial\omega} & = & -p_{f}(p^{\ast}G\epsilon)=-\frac{G}{\omega},\\
\frac{\partial F(G,p^{\ast},\omega,\epsilon)}{\partial\epsilon} & = & -\omega Gp^{\ast}p_{f}^{\prime}(p^{\ast}G\epsilon)=-\frac{G}{\epsilon}.
\end{eqnarray}
 Applying the implicit function theorem yields the set of derivatives
of $p^{\ast}$, 

\begin{eqnarray}
\frac{\partial p^{\ast}(G,\omega,\epsilon)}{\partial G} & = & -\left(\frac{\partial F(G,p^{\ast},\omega,\epsilon)}{\partial p^{\ast}}\right)^{-1}\cdot\frac{\partial F(G,p^{\ast},\omega,\epsilon)}{\partial G}\nonumber \\
 & = & 0,\\
\frac{\partial p^{\ast}(G,\omega,\epsilon)}{\partial\omega} & = & -\left(\frac{\partial F(G,p^{\ast},\omega,\epsilon)}{\partial p^{\ast}}\right)^{-1}\cdot\frac{\partial F(G,p^{\ast},\omega,\epsilon)}{\partial\omega}\nonumber \\
 & = & -\frac{p^{\ast}(G,\omega,\epsilon)}{\omega},\\
\frac{\partial p^{\ast}(G,\omega,\epsilon)}{\partial\epsilon} & = & -\left(\frac{\partial F(G,p^{\ast},\omega,\epsilon)}{\partial p^{\ast}}\right)^{-1}\cdot\frac{\partial F(G,p^{\ast},\omega,\epsilon)}{\partial\epsilon}\nonumber \\
 & = & -\frac{p^{\ast}(G,\omega,\epsilon)}{\epsilon},
\end{eqnarray}
which are solved by
\begin{equation}
p_{L}^{\ast}:=p^{\ast}=\frac{1}{\lambda\epsilon\omega},\label{eq:psimplestar}
\end{equation}
where $\lambda$ is a constant independent of $\omega$ and $\epsilon$.
We note that we did not make explicit assumptions on the distribution
of membrane potentials $P_{V}(V)$, which is determined by the setup
of the external network, i.e.~the external input current $I^{0}$,
the coupling strengths $\epsilon^{\text{ext,ex}}$ and $\epsilon^{\text{ext,in}}$
as well as the firing rates $\nu^{\text{ext,ex}}$ and $\nu^{\text{ext,in}}$.
With a different, lengthier approach based on a second order expansion
of $p_{f}$ one can derive an analytical estimate of $\lambda$ \cite{inPrep}.
\textcolor{black}{Fig.~\ref{fig:criticalconn} displays this analytical
approximation for $p_{\text{L}}^{\ast}$ and its agreement with numerical
simulations. For connectivity larger than $p_{\text{L}}^{\ast}$ there
is stable propagation of synchrony even in networks with linear dendritic
interactions and the size of the propagating pulse fluctuates around
the stable fixed point $G_{2}$ of Eq.~\eqref{eq:aveGroupSize}.
(For very large connectivity pathological high-frequency spiking activity
can emerge due to spontaneous chain activation.)}

\subsection{Feed-forward chains with nonlinear coupling}

\begin{figure*}
\centering{}\includegraphics[width=1\textwidth]{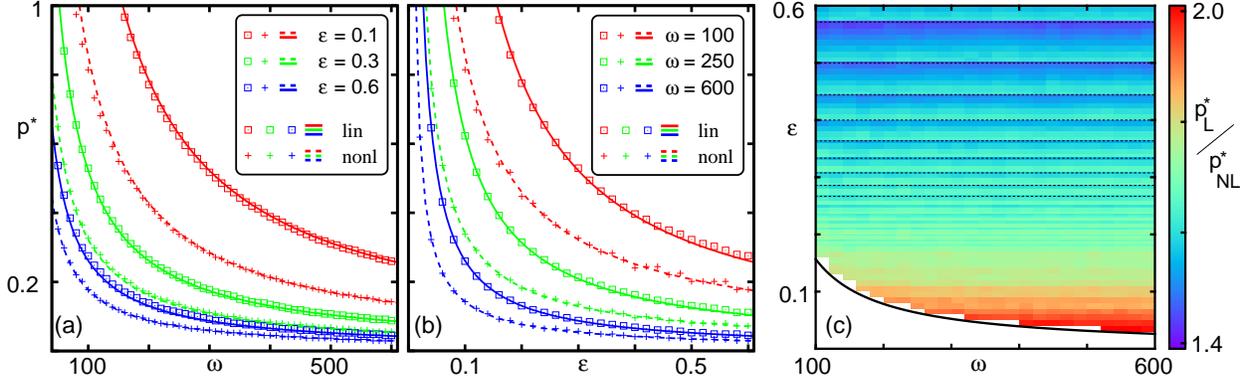}

\caption{(color online) Critical connectivity in isolated FFNs. (a,b) Network
simulations (symbols) agree well with analytical predictions (lines)
\eqref{eq:psimplestar} and \eqref{eq:pstarNLexopand}. The critical
connectivity decays with layer size and coupling strength. (c) The
reduction factor $c=p_{\text{L}}^{\ast}/p_{\text{NL}}^{\ast}>1$ shows
that nonlinear dendritic interaction compensates for reduced connectivity.
In both scenarios, with linear and nonlinear coupling, we find that
$p^{\ast}\propto\omega^{-1}$, such that the reduction factor is independent
of the layer size. In networks with linear couplings the critical
connectivity is $p^{\ast}\propto\epsilon^{-1}$, whereas in networks
with nonlinear coupling the dependence on $\epsilon^{-1}$ is nonlinear.
Therefore the reduction factor increases with decreasing coupling
strength. Dashed horizontal lines indicate jumps in the reduction
factor where the number of inputs needed for dendritic spike generation
changes. \label{fig:criticalconn}}
\end{figure*}
We now consider networks incorporating nonlinear dendritic interactions
and show that and how the connectivity and number of active neurons
required for propagation of synchrony is smaller. In such networks,
the mechanism underlying propagation of synchrony is different, because
it is supported predominantly by nonlinearly enhanced inputs. This
implies that the maximal input is bounded by $\kappa$, leading to
a saturation in the return map \eqref{eq:aveGroupSize} (cf.~Fig.~\ref{fig:averagegroup}c).
The saturation enables propagating pulses of a size substantially
smaller than $\omega$, in contrast to linearly coupled networks.
The discontinuity in the modulation function $\sigma$ induces a discontinuity
in $p_{f}\left(\sigma\left(x\right)\right)$ which prevents our previous
analytical method. We thus determine the critical connectivity by
a self-consistency approach. When a synchronous pulse arrives at a
specific layer, the summed excitatory input strength $x$ is either
smaller or larger than the dendritic threshold $\Theta_{b}$. For
sufficiently small $\Theta_{b}$, the spiking probability of a neuron
due to a subthreshold input is much smaller than due to a suprathreshold
input, i.e.~$p_{f}(\Theta_{b})\ll p_{f}(\kappa)$. Thus only a small
fraction of neurons receive an input smaller than $\Theta_{b}$ and
is elicited to spike. We approximate $p_{f}(x)=0$ for $x\leq\Theta_{b}$.
When there is persistent propagation of synchrony, $p_{\gamma}$,
which denotes the fraction of neurons that receive sufficiently strong
input to reach the dendritic threshold, is constant throughout the
layers. The total spiking probability of a single neuron upon the
arrival of the synchronous pulse is then given by the product $p_{\gamma}p_{f}(\kappa)$.
The probability 
\begin{equation}
p^{\#}(g)=\binom{\omega}{g}\left(p_{\gamma}p_{f}(\kappa)\right)^{g}\left(p_{\gamma}p_{f}(\kappa)\right)^{\omega-g}
\end{equation}
for $g$ neurons to spike synchronously follows a binomial distribution.
By combining the total spiking probability and the topological connection
probability $p$, we compute the probability

\begin{eqnarray}
P\left(k\right) & = & \sum_{g=k}^{\omega}\binom{g}{k}p^{k}(1-p)^{g-k}p^{\#}(g)\label{eq:Pinp}\\
 & = & \binom{\omega}{k}\left(p_{\gamma}pp_{f}(\kappa)\right)^{k}\left(1-p_{\gamma}pp_{f}(\kappa)\right)^{\omega-k}\label{eq:Pinp2}
\end{eqnarray}
that a neuron of the subsequent layer receives exactly $k$ synchronous
spikes. Thus, $k$ itself is binomially distributed and we denote
its mean value by $\delta$ and its standard deviation by $\sigma_{\delta}$.
Using a Gaussian approximation of the Binomial distribution yields
the self-consistent equation 
\begin{eqnarray}
p_{\gamma} & = & \sum_{k=\left\lceil \Theta_{b}/\epsilon\right\rceil }^{\omega}P(k)\\
 & \approx & \int_{\frac{\Theta_{b}}{\epsilon}}^{\infty}\frac{1}{\sqrt{2\pi}\sigma_{\delta}}\exp\left(-\frac{1}{2}\left(\frac{k-\delta}{\sigma_{\delta}}\right)^{2}\right)dk\\
 & = & \frac{1}{2}\left(1-\mbox{Erf}\left[\frac{\frac{\Theta_{b}}{\epsilon}-\delta}{\sqrt{2}\sigma_{\delta}}\right]\right)\\
 & =: & \frac{1}{2}\left(1+\mbox{Erf}\left(\frac{n}{\sqrt{2}}\right)\right),\label{eq:Defpg}
\end{eqnarray}
where we defined 
\begin{eqnarray}
n & := & \frac{\delta-\Theta_{b}/\epsilon}{\sigma_{\delta}}\label{eq:defn}\\
 & = & \frac{\omega\cdot p_{\gamma}\cdot p\cdot p_{f}(\kappa)-\Theta_{b}/\epsilon}{\sqrt{\omega\cdot p_{\gamma}\cdot p_{f}(\kappa)\cdot p\cdot(1-p_{\gamma}\cdot p_{f}(\kappa)\cdot p)}}
\end{eqnarray}
as the distance between the average number of inputs and the number
needed to reach the onset of the nonlinearity, measured in units of
$\sigma_{\delta}$. Solving definition \eqref{eq:defn} for $p$ that
occurs as an argument of $\delta$ and $\sigma_{\delta}$ and using
Eq.~\eqref{eq:Defpg} yields the connection probability in terms
of $n$, 
\begin{equation}
p_{\text{NL}}=\frac{n^{2}\epsilon+2\text{\ensuremath{\Theta_{b}}}+n\sqrt{n^{2}\epsilon^{2}+4\Theta_{b}\left(\epsilon-\frac{\text{\ensuremath{\Theta_{b}}}}{\omega}\right)}}{p_{f}(\kappa)\epsilon(n^{2}+\omega)\left(1+\mbox{Erf}\left(\frac{n}{\sqrt{2}}\right)\right)}.\label{eq:pstarNLfull}
\end{equation}
 For a certain setup of the FFN with variable connectivity, $p_{\text{NL}}\left(n\right)$
is the connectivity for which a stationary propagation of synchrony
occurs with a certain $n$. Any $p_{\text{NL}}$ above the critical
connectivity $p_{\text{NL}}^{\ast}$, has two preimages $n$, corresponding
to the group sizes $G_{1}$ and $G_{2}$, $p_{\text{NL}}^{\ast}$
has one preimage, and any $p_{\text{NL}}$ below $p_{\text{NL}}^{\ast}$
has none, cf.~Fig.~\ref{fig:averagegroup}c. Thus, $p_{\text{NL}}(n)$
has one global minimum at $n=n^{\ast}$ where $\left.\frac{dp_{\text{NL}}(n)}{dn}\right|_{n=n^{\ast}}=0$
and the critical connectivity is $p_{\text{NL}}(n^{\ast})=p_{\text{NL}}^{\ast}$.

The comparison of the results for linearly and nonlinearly coupled
FFNs is particularly enlightening in the limit of large layer size
($\omega\gg1$) and small coupling strengths ($\epsilon\ll\Theta-V^{\textsf{reset}}$).
We fix the maximal input to a neuron from the previous layer, $\epsilon\omega=\text{const}$,
to preserve the network state and expand Eq.~\eqref{eq:pstarNLfull}
in a power series around $\omega\rightarrow\infty$ and $\epsilon\rightarrow0$.
Considering the leading terms we find
\begin{equation}
p_{\text{NL}}\approx2\frac{\Theta_{b}+n\sqrt{\Theta_{b}\left(\epsilon-\frac{\Theta_{b}}{\omega}\right)}}{p_{f}\left(\kappa\right)\epsilon\omega\left(1+\mbox{Erf}\left(\frac{n}{\sqrt{2}}\right)\right)}.\label{SUB:pnlapp}
\end{equation}
We note that propagation of synchrony mediated by dendritic spikes
is enabled if a sufficiently large fraction of neurons of each layer
receives a total input larger or equal $\Theta_{b}$; this implies
in particular $\Theta_{b}<\omega\epsilon$. Moreover, if the connectivity
within the FFN is low, stable propagation even requires $\Theta_{b}\ll\omega\epsilon$
and $p_{\text{NL}}$ further simplifies to
\begin{equation}
p_{\text{NL}}\approx\frac{2\Theta_{b}}{p_{f}(\kappa)\epsilon\omega}\frac{1+n\sqrt{\frac{\epsilon}{\Theta_{b}}}}{1+\text{Erf}\left(\frac{n}{\sqrt{2}}\right)}.\label{SUB:pnlapp2}
\end{equation}
As described above, the critical connectivity is given by the minimum
of $p_{\text{NL}}$ as a function of $n$, which is assumed at $n=n^{*}$.
$\left.\frac{dp_{\text{NL}}(n)}{dn}\right|_{n=n^{\ast}}=0$ yields
$n^{*}$ as an implicit function of $\frac{\Theta_{b}}{\epsilon}$,
\begin{equation}
\sqrt{\frac{\Theta_{b}}{\epsilon}}=\sqrt{\frac{\pi}{2}}\exp\left(\frac{n^{\ast^{2}}}{2}\right)\left(1+\mbox{Erf}\left(\frac{n^{\ast}}{\sqrt{2}}\right)\right)-n^{\ast}.\label{SUB:impliciten}
\end{equation}
For better readability, we define
\begin{eqnarray}
\beta\left(\frac{\Theta_{b}}{\epsilon}\right): & = & \frac{1}{2}\left(1+\text{Erf}\left[\frac{n^{\ast}}{\sqrt{2}}\right]\right)-n^{\ast}\frac{e^{-\frac{n^{\ast^{2}}}{2}}}{\sqrt{2\pi}},\label{SUB:defbeta}
\end{eqnarray}
 where $n^{*}=n^{*}\left(\frac{\Theta_{b}}{\epsilon}\right)$ as given
by Eq.~\eqref{SUB:impliciten}. Combining Eqs.~(\ref{SUB:pnlapp2}-\ref{SUB:defbeta})
enables to simplify the critical connectivity to

\begin{equation}
p_{\text{NL}}^{\ast}=\frac{\Theta_{b}}{p_{f}(\kappa)\epsilon\omega}\cdot\frac{1}{\beta\left(\Theta_{b}/\epsilon\right)},\label{eq:pstarNLexopand}
\end{equation}
which depends nonlinearly on the number of spikes needed to reach
the dendritic threshold ($\Theta_{b}/\epsilon$) through the function
$1/\beta\left(\cdot\right)$. One can show that $\beta(\Theta_{b}/\epsilon)$
increases with decreasing coupling strength $\epsilon$ from $\beta(\Theta_{b}/\epsilon)=0.5$
for large $\epsilon$ and becomes maximal in the limit of small couplings,
$\lim_{\epsilon\rightarrow0}\beta(\Theta_{b}/\epsilon)=1$. Fig.~\ref{fig:criticalconn}
displays the results for $p_{\text{NL}}^{\ast}$ together with numerical
simulations. As in the linearly coupled network the critical connectivity
decays with layer size and coupling strength, but the dependence on
$1/\epsilon$ is nonlinear. The factor

\begin{equation}
c:=\frac{p_{\text{L}}^{\ast}}{p_{\text{NL}}^{\ast}}=\frac{p_{f}(\kappa)}{\lambda\Theta_{b}}\cdot\beta\left(\frac{\Theta_{b}}{\epsilon}\right),\label{eq:defofc}
\end{equation}
 by which the nonlinear dendritic interactions reduce the required
network connectivity increases with decreasing threshold $\Theta_{b}$
and increasing enhancement $\kappa$. Fig.~\ref{fig:criticalconn}c
illustrates the numerically obtained reduction of connectivity: The
critical connectivity $p_{\text{NL}}^{\ast}$ is smaller over the
whole parameter range; the reduction is most effective for small $\epsilon$
and largely independent of $\omega$.

Nonlinear dendrites thus foster propagation of synchrony. We remark
that our model still overestimates the capability of linearly coupled
networks to propagate synchrony: Upon synchronous input linearly coupled
groups of neurons generate synchronous output (if they generate output
at all). This is a consequence of the infinitesimally short synaptic
currents. In neurons with extended synaptic currents the timing of
the output strongly depends on the neurons' state and input strength.
In contrast the timing of somatic action potentials elicited by dendritic
spikes is largely independent thereof. We therefore expect the effect
of nonlinear dendrites to be even stronger in networks of biologically
more detailed neurons as considered in the next section.

\subsection{Recurrent networks}

\begin{figure*}
\centering{}\includegraphics[width=1\textwidth]{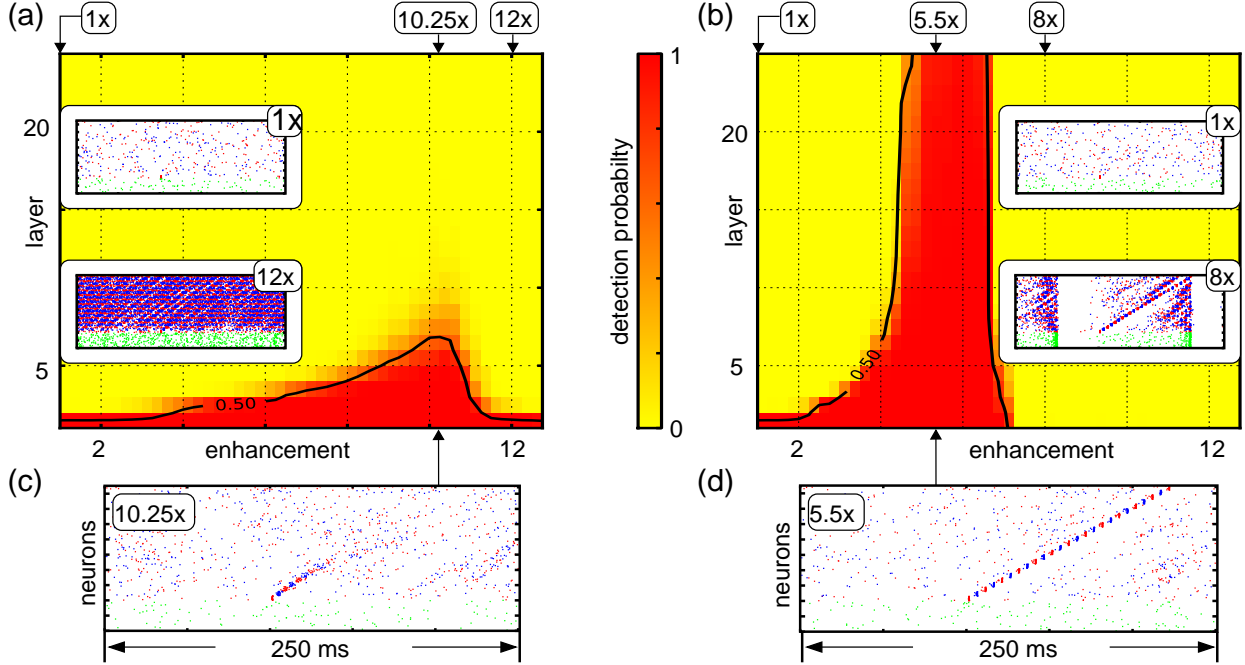}

\caption{(color online) Nonlinear dendritic interactions enable guided propagation
of synchrony in random networks ($N=10.000$, $p=0.03$, $\omega=30$).
(a,b): Detection probability of induced synchronous activity in conventional
networks (a) and networks with nonlinear dendrites (b). Insets of
(a,b), and (c,d): Typical examples of network activity for different
synaptic enhancement factors (red and blue: spikes of neurons within
the FFN, green: spikes of neurons randomly sampled from the remaining
neurons). \label{fig:network}}
\end{figure*}

The main findings generalize in two ways, to FFNs occurring in recurrent
random networks and to biologically more detailed models. For such
systems, we show that in nonlinearly coupled networks stable propagation
naturally emerges where it is difficult to achieve in linearly coupled
networks. In contrast to isolated FFNs studied above, we now account
for effects of the FFN on the surrounding network and its feedback.
Further, we choose a more detailed neuron model (see section ``Models
and Methods'') to assure that main assumptions underlying the analytically
tractable model are not crucial for stable propagation of synchrony.
In particular, we show that systems with temporally extended postsynaptic
responses and temporally extended nonlinear dendritic interaction
window exhibit qualitatively the same phenomena as found above.

We consider networks of randomly connected conductance-based leaky
integrate-and-fire neurons (cf.~Eqs.~\eqref{eq:CBModel}-\eqref{eq:DSModel}).
The networks consist of $N^{\text{E}}$ excitatory and $N^{\text{I}}$
inhibitory neurons. A directed connection between two neurons is present
with probability $p$. As for the isolated FFNs considered above,
we construct the network such that the ground state in the absence
of synchronous activity is characterized by balanced excitatory and
inhibitory input, which results in an asynchronous irregular spiking
activity. For simplicity, all neurons have the same parameters, e.g
$C_{l}^{\text{m}}=C^{\text{m}}$, $g_{l}^{L}=g^{L}$, etc. 

First we set up a model network, where the total excitation and inhibition
to the neurons is balanced such that the spiking activity is asynchronous
irregular. The external constant current $I^{0}$ determines together
with the leak conductance $g^{L}$ and the resting potential $V^{\text{rest}}$
the asymptotic membrane potential in the absence of incoming spikes,
\begin{equation}
V^{\infty}=V^{\text{rest}}+\frac{I^{0}}{g^{L}}.
\end{equation}
Additionally, each neuron receives excitatory and inhibitory random
Poissonian spike trains. The frequencies are denoted by $\nu^{\text{ext,ex}}$
and $\nu^{\text{ext,in}}$ and the ratio between them is chosen such
that it equals the ratio of the number of excitatory and inhibitory
neurons in the network,
\begin{equation}
\frac{\nu^{\text{ext,ex}}}{\nu^{\text{ext,in}}}=\frac{N^{\text{E}}}{N^{\text{I}}}.
\end{equation}
This ensures that each neuron receives the same ratio of excitatory
and inhibitory input from both the network and external sources when
neurons in the excitatory and inhibitory network populations spike
on average with the same mean rate. All excitatory as well as all
inhibitory connections have the same strength, i.e. $g_{lj}^{\text{max}}=g^{\text{ex}}$
for excitatory and $g_{lj}^{\text{max}}=g^{\text{in}}$ for inhibitory
connections. The ratio of the peak postsynaptic potentials due to
an inhibitory and an excitatory input at the asymptotic membrane potential
$V^{\infty}$ is approximately given by 
\begin{equation}
g^{\text{rat}}:=\frac{g^{\text{in}}\left|V^{\infty}-E^{\text{in}}\right|}{g^{\text{ex}}\left|V^{\infty}-E^{\text{ex}}\right|}.
\end{equation}
 We set

\begin{eqnarray}
g^{\text{rat}} & \stackrel{!}{=} & \frac{N^{\text{E}}}{N^{\text{I}}}=\frac{\nu^{\text{ext,ex}}}{\nu^{\text{ext,in}}}\\
g^{\text{in}} & = & \frac{\left|V^{\infty}-E^{\text{ex}}\right|}{\left|V^{\infty}-E^{\text{in}}\right|}\cdot\frac{N^{\text{E}}}{N^{\text{I}}}\cdot g^{\text{ex}}\label{SUP:SynScale}
\end{eqnarray}
 to obtain balanced activity.

In contrast to the model considered in the first sections, now excitatory
neurons have a non-zero time window $\Delta t$ for nonlinear dendritic
modulation. When the strength of the excitatory input within $\Delta t$
exceeds a threshold, a current pulse is injected into the soma, modeling
the effect of a dendritic spike. The neuron parameters for this phenomenological
model are chosen according to experimental findings to reproduce the
time course of the membrane potential in response to a dendritic spike
quantitatively (see section ``Models and Methods'').

Considering a random network we detect naturally occurring weak feed-forward-structures
suitable for signal transmission in the following way: We randomly
choose a group of $x$ neurons to be the first layer. The second layer
is composed of $x$ neurons out of those receiving the largest numbers
of connections from the initial group. By repeating this selection
process $l$ times, we identify a FFN consisting of $l$ layers. In
each selection step, we exclude the $x$ neurons of the previous layer,
but allow to choose neurons which are members of the layers preceding
the previous one. The high-connectivity subnetwork selected from an
existing random network as described above by construction has a slightly
higher than average connection probability. Therefore this structure
is particularly well suited to enable propagation of synchrony. Alternatively
one can assign neurons randomly to the different layers and compensate
for smaller connectivity by, e.g., larger layer sizes according to
Eq.~\eqref{eq:pstarNLexopand}.

The measurements start after an equilibration phase (initially the
network is at rest). In the ground state, the network generates balanced
irregular activity. Propagation of synchrony is initiated by exciting
the neurons of the first layer to spike within a short time interval,
which is smaller than the time window of dendritic integration, $\Delta t$.
This leads to an increased input to the second layer after a delay
time $\tau$. This, in turn, may lead to highly synchronous spiking
of a certain number of neurons of the second layer (possibly supported
by dendritic spikes) and therewith to synchronous spiking after another
delay time $\tau$ in the third layer etc. Propagation of synchrony
requires that (i) the total input of a layer to its successor within
the FFN is sufficiently strong and (ii) that the input to the remaining
network is sufficiently weak to avoid excitation of too many neurons
to synchronous spiking. Requirement (ii) prevents pathological activity
such as ``synfire-explosions'' \cite{EmbSyns}. 

After initiating a propagation of synchrony by exciting the neurons
of the first group to spike within a short time interval, we measure
the probability of detecting a synchronous pulse in the subsequent
groups (see section ``Models and Methods''), cf.~Fig.~\ref{fig:network}a,b.
Although the average connectivity within the identified FFN is significantly
larger than the overall connectivity $p$, it is still small and propagation
of synchronous activity is very unlikely (upper insets of Fig.~\ref{fig:network}a,b).
We find that it is not sufficient to choose high-connectivity subnetworks
as FFNs (as described above) to obtain a stable propagation of synchrony,
but that the synapses within the FFN have to be strengthened. To study
the transition to propagation we strengthen the synapses within the
FFN gradually. As suggested by the results on isolated chains, we
observe a propagation of synchrony over more and more layers for moderate
enhancements (Fig.~\ref{fig:network}c,d). For very strong enhancements
the feedback from the network becomes important: The synaptic amplification
leads to an increased spontaneous activity within the FFN and this
in turn results in an increased background activity. The overall increased
spiking activity causes spontaneous synchronous pulses and a separation
of the induced synchronous signal from the background activity is
not possible anymore (the detection probability decreases, see Fig.~\ref{fig:network}a,b
and lower insets).

In agreement with previous studies (cf.~\cite{Vogels2005}), we find
that in the linearly coupled networks considered a synchronous pulse
propagates only over a few layers, even in the optimal enhancement
range (Fig.~\ref{fig:network}a). In contrast, networks incorporating
nonlinear dendrites support stable propagation of synchrony (Fig.~\ref{fig:network}b)
in a substantial region of parameter space. In addition the propagation
is enabled for enhancements considerably smaller than the optimal
enhancement for networks with linear dendrites.

\section{Discussion}

In conclusion, we have analyzed strongly diluted networks with linear
and nonlinear dendritic interactions. We have shown that and how nonlinear
dendritic interactions may enhance and stabilize synchrony propagation
in both isolated feed-forward chains and in recurrent network structures.
Moreover, our results show that such local nonlinear interactions
support the separation of propagating synchrony and asynchronous background
activity. Earlier works \cite{EmbSyns,Vogels2005} did not take into
account supralinear amplification of synchronous activity. There is
one study \cite{Vogels2005} using existing connections in recurrent
networks to create diluted chains assuming strongly enhanced synapses
and at the same time specifically tuned neuron properties; still synchrony
could propagate only over a few groups. In contrast, the results presented
above indicate that a reliable propagation is achieved by only mildly
adapted synapses and without specifically tuning or changing neuron
properties or rewiring the network.

In a recent study \cite{MemmTimme2012} incorporating nonlinear dendrites
it has been shown that synchronous activity can propagate in purely
random networks without modified connections. There are no specific
propagation paths but neurons are recruited in a quasi-random manner.
Our results above now indicate that specific feed-forward chains that
naturally occur in random neural circuits are capable of persistently
propagating synchronous signals if their synaptic strengths are increased.
The strengths required in the presence of nonlinear interactions are
common in biological neural circuits \cite{chexp} and may well be
generated by learning, e.g. through spike-timing-dependent plasticity.

Dendritic (coupling) nonlinearities therefore offer a viable mechanism
for guiding synchrony through weakly structured random topologies.

Recently \cite{LJF2010} feed-forward chains with slow dendritic (probably
calcium) spikes have been simulated to check the possibility of the
occurrence of specific spike patterns that are experimentally observed
in the higher vocal center of song birds. Our theoretical work now
yields analytic insights about the collective dynamics of circuits
with fast dendritic (sodium) spikes. Fast dendritic spikes have been
found in the hippocampus and in the neocortex and may thus be involved
in hippocampal replay, memory formation and other computational processes.
Experimentally, the influence of fast dendritic spikes could be directly
checked by selectively blocking dendritic sodium channels (e.g. \cite{block}
indicates that the types of sodium channels in the dendrite and soma
are different) and thereby distinguishing effects coming from non-additive
coupling via fast dendritic spikes from those induced by other mechanisms.
During the last decade, the number of neurons simultaneously accessible
has multiplied from a few to the order of $10^{2}$ neurons, with
this rapid trend further ongoing. When recording the activity of a
substantial fraction of neurons of a local circuit synchrony propagation
should be clearly detectable and analyzable. Our results suggest that
synchrony propagation and thus spike patterns should be influenced
if dendritic sodium channels and thus fast dendritic spikes are blocked.
Specifically, in the hippocampus, the precision of (replayed) spike
patterns will decrease or the patterns vanish after blocking. Such
experiments would thus provide a direct test of how non-additive coupling
is exploited for the collective dynamics of neural circuits. Once
the connectome, i.e. the structural synaptic connectivity, of neural
circuits becomes available in the future \cite{connect}, the relative
impact of synaptic, structural to dynamic features of single neurons
on circuit dynamics may be well distinguishable.

The basis model of pulse-coupled units considered here is applicable
to a range of systems in nature, not only neural circuits but also,
e.g., earthquakes emerging from abruptly relaxing tectonic plates,
and fireflies interacting by exchanging light flashes (e.g.\ \cite{MS90HH95}).
We have now studied the impact of nonlinear input modulation on collective
network dynamics and derived methods for their analysis that may be
useful also in a non-neuronal setting. Interestingly, very recent
results \cite{MC10} have shown that fireflies are more prone to respond
to synchronous flashes rather than to asynchronous ones suggesting
a direct application of our model.

\section{Acknowledgments}

Supported by the BMBF (grant no. 01GQ1005B), the DFG (grant no. TI
629/3-1) and the Swartz Foundation. Simulation results were partly
obtained using the simulation software NEST \cite{Gewaltig2007}.

\section{Appendix}

\subsection{Parameters for Figs.~\ref{fig:averagegroup} and \ref{fig:criticalconn}}

The single neuron parameters and the coupling delay are $\tau^{\text{m}}=1/\gamma=14$ms,
$\Theta=15$mV, $V^{\text{reset}}=0$mV, $t^{\text{ref}}=2$ms and
$\tau=10$ms \cite{AMABK07,DA01}. The external input is characterized
by $\epsilon^{\text{ext,ex}}=-\epsilon^{\text{ext,in}}=0.5$mV \cite{chexp,MW86DT96},
$\nu^{\text{ext}}=\nu^{\text{ext,ex}}=\nu^{\text{ext,in}}=3$kHz and
$V^{\infty}=5$mV. The parameters of the dendritic modulation function
were chosen according to single neuron measurements as $\Theta_{b}=4$mV
and $\kappa=11$mV \cite{DSpikes}.

The maps and transition matrices presented in Fig.~\ref{fig:averagegroup}
are derived for $\omega=100$ and $\epsilon_{ij}=\epsilon=0.3\text{mV}$.
To obtain the distribution of active neurons $g_{i+1}$ in layer $i+1$,
we excite $g_{i}$ neurons of the first layer to spike simultaneously
and measure the number of active neurons in the following layer. For
each value of $g_{i}$ we calculate the distribution for $m=1000$
different realizations of the FFN and initial conditions.

In Fig.~\ref{fig:criticalconn}, existing connections within the
FFN have strengths $\epsilon_{ij}=\epsilon$. We determine the critical
connectivity for $\epsilon\in\left[0.05\text{mV},0.6\text{mV}\right]$
and layer sizes $\omega\in\left[50,600\right]$ as follows: We construct
a FFN consisting of $20$ layers, with $\omega$ neurons in each layer
and connect neurons of successive layers with probability $p\in\left[0,1\right]$.
After an equilibration time $t^{\text{init}}$ (initially the network
is at rest), we initiate propagation of synchrony by exciting all
neurons of the first layer to spike simultaneously. We then check
whether the synchronous pulse propagates up to layer $i$, i.e. whether
there is synchronous activity in layer $i$ at time $t_{i}^{\text{exp}}=t^{\text{init}}+(i-1)\cdot\tau.$
We consider the propagation for a certain setup specified by $\epsilon,$
$\omega$ and $p$ as `successful', if a synchronous pulse propagates
along the whole FFN in more than $50\%$ of $o=31$ realizations of
the FFN with different initial conditions. We derive the critical
connectivities $p_{\text{L}}^{\ast}$ and $p_{\text{NL}}^{\ast}$
up to a resolution of $\frac{\Delta p}{p}=5\cdot10^{-3}$ by repeatedly
bisecting the interval $[0,1]$ and testing the success of propagation.

\subsection{Parameters for Fig.~\ref{fig:network}}

For the network simulations, we employed the simulation software NEST
\cite{Gewaltig2007}, by the NEST Initiative, available at www.nest-initiative.org.
The networks had a total number of $N=10.000$ neurons with $N^{\text{E}}=8.000$
and $N^{\text{I}}=2.000$. For simplicity all neurons are considered
identical, i.e. $C_{l}^{\text{m}}=C^{\text{m}}$, $g_{l}^{L}=g^{L}$,
$V_{l}^{\text{rest}}=V^{\text{rest}}$, $I_{l}^{0}=I^{0}$, $\Theta_{l}=\Theta$,
$t_{l}^{\text{ref}}=t^{\text{ref}}$ and $V_{l}^{\text{reset}}=V^{\text{reset}}$.
The single neuron parameters are $C^{\text{m}}=400$pF, $V^{\text{rest}}=V^{\text{reset}}=-65$mV,
$g^{L}=25$nS, $\Theta=-50$mV, $t^{\text{ref}}=3$ms \cite{AMABK07,SJTYS00},
$I^{0}=250$pA, and the frequencies of the external inputs are $\nu^{\text{ext,ex}}=2.4$kHz
and $\nu^{\text{ext,in}}=0.6$kHz. The recurrent connectivity in cortical
and hippocampal networks is sparse: Connection probabilities between
$1\%$ and $10\%$, depending on the distance and the region have
been estimated (e.g.~\cite{AMABK07,chexp,MW86DT96}), for our simulations
we choose $p=0.03$.

The time constants of the excitatory (AMPA) conductances are $\tau^{\text{A,1}}=2.5$ms
and $\tau^{\text{A,2}}=0.5$ms \cite{JMS93LT95}. For simplicity,
we choose the same time constants for the inhibitory (GABA$_{\text{A}}$)
conductances, $\tau^{\text{G,1}}=2.5$ms and $\tau^{\text{G,2}}=0.5$ms.
The reversal potentials are $E^{\text{ex}}=0$mV and $E^{\text{in}}=-75$mV
\cite{DA01,AMABK07}. The strengths of experimentally observed pEPSPs
due to single inputs range from small values like $0.1$mV to larger
values like $2$mV \cite{AMABK07,chexp,MW86DT96}. For non-enhanced
couplings, we set $g^{\text{ex}}=0.6$nS, which corresponds to a pEPSP
of approximately $0.3$mV at rest. According to Eq.~\eqref{SUP:SynScale},
the coupling strength of the inhibitory synapses are $g^{\text{in}}=-6.6$nS
to maintain balanced input. This configuration results in an asynchronous
irregular ground state with a spontaneous firing rate $\nu\approx1.8$Hz.

The parameters of the dendritic spike current are chosen according
to single neuron measurements in hippocampal cells: $\Delta t=2$ms
\cite{DSpikes}, $g^{\Theta}=8.65$nS (corresponding to a pEPSP of
about $3.8$mV at rest \cite{DSpikes}), $\tau^{\text{DS}}=2.7$ms
(such that $\tau+\tau_{DS}=4.7$ms and the peak of the depolarization
is reached approximately $5$ms after presynaptic spiking), $A=55$nA,
\textbf{$B=64$}nA, $C=9$nA, $\tau^{\text{DS,1}}=0.2$ms, $\tau^{\text{DS,2}}=0.3$ms,
$\tau^{\text{DS,3}}=0.7$ms and $t^{\text{ref,DS}}=5.2$ms. The correction
factor, which modulates the strength of the dendritic spike, is found
by fitting a linear correction function, $c(g)=\max\left\{ 1.5-g\cdot0.053\text{nS}^{-1},0\right\} $,
such that the experimentally observed region of saturation is obtained.
The dynamics of the neuron model incorporating the mechanism for dendritic
spike generation is illustrated in Fig.~\ref{fig:props}.

For calculating the SNR we use an $a=0.99$ and \textbf{$b=2$ }and
an expected width of the synchronous pulse $t^{w}=10$ms; the result
is insensitive to changes in these parameters. The expected interval
between successive synchronous active layers,$\Delta t^{\text{exp}}$,
is chosen from the interval $\left[2\text{ms},7\mbox{\text{ms}}\right]$
such that the signal, $\sum_{i}S_{i}$, is maximized (cf.~section
``Models and Methods''). The control interval time interval for
the estimation of the noise level is $\Delta t^{obs}=15$s. The detection
probability shown in Fig.~\ref{fig:network}a,b is the fraction of
successful propagations obtained from $10$ different network realizations,
where for each network setup propagation of synchrony was tested for
$20$ initial conditions.

All measurements start after an initial equilibrium phase of $t{}^{0}=4000$ms.

\end{document}